# The Relation between the Optical Variability Timescale, Magnetic Field of Jets, and Black Hole Spin in Active Galactic Nuclei

Yongyun Chen (陈永云)[1], Qiusheng Gu (顾秋生)[2], Junhui Fan (樊军辉)[3], Dingrong Xiong (熊定荣)[4], Xiaoling Yu (俞效龄)[1],  
Xiaotong Guo (郭晓通)[5], Nan Ding (丁楠)[6], and Ting-Feng Yi (易庭丰)[7]  
[1] College of Physics and Electronic Engineering, Qujing Normal University, Qujing 655011, People's Republic of China; ynkmcyy@yeah.net  
[2] School of Astronomy and Space Science, Nanjing University, Nanjing 210093, People's Republic of China; qsgu@nju.edu.cn  
[3] Center for Astrophysics, Guang zhou University, Guang zhou510006, People's Republic of China  
[4] Yunnan Observatories, Chinese Academy of Sciences, Kunming 650011, People's Republic of China  
[5] Anqing Normal University, 246133, People's Republic of China  
[6] School of Physical Science and Technology, Kunming University 650214, People's Republic of China  
[7] Department of Physics, Yunnan Normal University, Kunming 650500, People's Republic of China  
*Received 2025 April 29; revised 2025 August 22; accepted 2025 September 16; published 2025 October 23*

## Abstract

We investigate the relationship among the jet magnetic field, black hole spin, black hole mass, Eddington ratio, and optical variability timescales in jetted active galactic nuclei (AGNs). By fitting a damped random walk (DRW) model to the *g*-band light curves, we obtain the characteristic variability timescale ($\tau_{\rm DRW}$) for 41 jetted AGNs with precise supermassive black hole (SMBH) mass measurements. Our main results are as follows: (i) Our analysis reveals a significant correlation between the jet magnetic field ($B_{\rm 1\,pc}$), black hole spin ($j$) and the characteristic variability timescale within our sample. These findings suggest that the optical variability of jetted AGNs is influenced by the jet magnetic field and black hole spin. Furthermore, the characteristic variability timescale aligns with the electron escape timescale, as evidenced by the relationship between the characteristic variability timescale and jet magnetic field ($\tau_{\rm DRW} \propto B_{\rm 1\,pc}^{0.76\pm0.22}$). (ii) We confirm a significant correlation between the characteristic variability timescale and SMBH mass, expressed as: $\log \tau_{\rm DRW} = 0.52(\pm 0.21)\log M_{\rm BH}/M_\odot - 3.12(\pm 1.90)$, with an intrinsic scatter of 0.08 dex. The slope of this relationship is consistent with that between the thermal timescale and black hole mass. Our results support the hypothesis that magnetorotational instability fluctuations drive the intrinsic variability observed in the light curves emitted by the AGN's accretion disk.

*Unified Astronomy Thesaurus concepts:* Active galactic nuclei (16); Galaxy jets (601); Supermassive black holes (1663); Time domain astronomy (2109); Time series analysis (1916)

## 1. Introduction

The optical variability exhibited by active galactic nuclei (AGNs) offers crucial insights into the physical processes occurring within the accretion disk of rapidly accreting supermassive black holes (SMBHs). These AGN accretion disks predominantly emit radiation in the rest-frame ultraviolet (UV) to optical wavelengths. Variability serves as a vital tool for studying AGNs; however, the origin of AGN variability remains ambiguous and is a subject of ongoing debate in the literature (e.g., J. Terrell 1967; A. P. Marscher & W. K. Gear 1985; S. J. Wagner & A. Witzel 1995; Y. E. Lyubarskii 1997; J. Kataoka et al. 2001; P. Uttley et al. 2003; S. G. Sergeev et al. 2005; B. Czerny 2006; I. M. McHardy et al. 2006; A. P. Marscher et al. 2008; M. Böttcher et al. 2010; R. F. Mushotzky et al. 2011; J. J. Ruan et al. 2012; C. M. Raiteri et al. 2017, 2024; M. Sun et al. 2020; G. Bhatta 2021; S. G. Jorstad et al. 2022). Over the past few decades, the study of optical variability in AGNs has witnessed remarkable progress. This advancement is largely attributable to the enhancement in the quality of light curves and the expansion of sample sizes (e.g., T. A. Matthews & A. R. Sandage 1963; U. Giveon et al. 1999; M. R. S. Hawkins 2002; D. E. Vanden Berk et al. 2004; W. H. de Vries et al. 2005; B. Sesar et al. 2006; A. Bauer et al. 2009; C. L. MacLeod et al. 2010, 2012; E. Morganson et al. 2014; Y.-H. Sun et al. 2014; X.-Y. Chen & J.-X. Wang 2015; V. P. Kasliwal et al. 2015; T. Simm et al. 2016; N. Caplar et al. 2017; Z. Li et al. 2018; P. Sánchez-Sáez et al. 2018; K. L. Smith et al. 2018; D. De Cicco et al. 2019; M. Laurenti et al. 2020; Y. Luo et al. 2020; Y. Tachibana et al. 2020; C. Xin et al. 2020; C. J. Burke et al. 2021; K. L. Suberlak et al. 2021; Z. Stone et al. 2022; W. Yu et al. 2022; X.-F. Hu et al. 2024; Z.-B. Su et al. 2024; T. Zhang et al. 2024; H. Zhang et al. 2024; H. Zhou & D. Lai 2024; P. Patel et al. 2025). A variety of scenarios have been put forward to elucidate the variability of AGNs. One such scenario involves the injection, acceleration, and cooling of particles within the jet, which may be disrupted by shock waves or turbulence (e.g., A. P. Marscher & W. K. Gear 1985; M. Sikora et al. 2001; G. Ghisellini et al. 2002; A. P. Marscher 2014). Another perspective focuses on geometric interpretations or alterations in jet regions (e.g., M. Villata & C. M. Raiteri 1999; A. P. Marscher et al. 2008; A. A. Abdo et al. 2010; C. M. Raiteri et al. 2017, 2024). Additionally, gravitational microlensing has been proposed as a possible cause (e.g., D. F. Torres et al. 2003), as well as kink instabilities and magnetic reconnection (e.g., N. Ding et al. 2019; S. G. Jorstad et al. 2022; X. Chang et al. 2024).

Intensive monitoring of nearby AGNs suggests that optical variability may be driven by X-ray variability, which illuminates the accretion disk and subsequently generates UV/optical variations (J. Clavel et al. 1992). This hypothesis can be directly tested using long and well-sampled light curves in X-ray, UV, and optical. According to this model, changes in X-ray emission propagate through the disk as an "echo," first modulating UV emission from the inner disk regions and later optical emission from the outer disk. The radial temperature profile of a standard disk predicts a time lag relationship between the variability of the X-ray source and the variability at other wavelengths, expressed as $\tau \propto \lambda^{4/3}$ (S. Collier et al. 1999). The X-ray variability includes short-term variability (high-frequency variability) and long-term variability (low-frequency variability; e.g., I. E. Papadakis et al. 2001; L. Mallick et al. 2018). Long-term X-ray variability may arise from the inward propagation of fluctuations in the local accretion rate (Y. E. Lyubarskii 1997), leading to corresponding changes in UV/optical radiation. These changes can further influence variations in the X-ray emission from the corona. Alternatively, long-term variability could result from variations in light-bending intensity caused by changes in the geometric position of the corona, a mechanism derived from the relativistic reflection scenario. In contrast, the physical origin of short-term X-ray variability remains uncertain but may be associated with turbulent processes within the corona, such as magnetic turbulence (e.g., magnetic turbulence; S. C. Noble & J. H. Krolik 2009).

However, observations over timescales of months or years indicate that the amplitude of UV/optical variability exceeds that of X-ray variability. This discrepancy implies that X-ray reprocessing alone cannot account for the full extent of UV/optical variability, necessitating intrinsic variability in UV–optical light curves emitted by AGN disks (J. H. Krolik et al. 1991; P. Arévalo et al. 2008, 2009; R. Edelson et al. 2015; P. Lira et al. 2015; J. V. Hernández Santisteban et al. 2020; J. M. M. Neustadt & C. S. Kochanek 2022; M. W. J. Beard et al. 2025). This variability may be caused by fluctuations in the UV–optical region of the accretion disk due to turbulence driven by magnetorotational instability (MRI; S. A. Balbus & J. F. Hawley 1991) or convection due to increased opacity in the ultraviolet optical area of the disk (Y.-F. Jiang & O. Blaes 2020). Using three-dimensional multifrequency radiation magnetohydrodynamic simulations, some authors also suggested that MRI turbulence is sufficient to produce intrinsic variability in UV light curves (A. Secunda et al. 2024, 2025).

Assuming that the optical variability stems from the accretion disk in close proximity to the supermassive black hole (SMBH), a range of models have been put forward to elucidate the observed variability patterns. Drawing upon the standard $\alpha$ disk model (N. I. Shakura & R. A. Sunyaev 1973), some scholars have suggested that the optical variability we observe may well be propelled by alterations in the mass accretion rate (N. A. Pereyra et al. 2006; S.-L. Li & X. Cao 2008; M. Kokubo 2015). From an observational standpoint, remarkable strides have been made in probing the physical origin of ultraviolet (UV)/optical variability. This has been achieved by delving into the correlation between optical variability and the inherent physical properties of AGNs. Notable results include the inverse correlation between variability amplitude and rest frame wavelength, as well as the inverse correlation between variability and luminosity and/or Eddington ratio (e.g., D. E. Vanden Berk et al. 2004; B. C. Wilhite et al. 2007; A. Bauer et al. 2009; C. L. MacLeod et al. 2010; T. Simm et al. 2016; N. Caplar et al. 2017; S. Rakshit & C. S. Stalin 2017). Many studies have also reported a correlation between variability and black hole mass (e.g, B. C. Wilhite et al. 2007, 2008; M. Wold et al. 2007; A. Bauer et al. 2009; B. C. Kelly et al. 2009; C. L. MacLeod et al. 2010; T. Simm et al. 2016; N. Caplar et al. 2017; C. J. Burke et al. 2021; H. Zhang et al. 2022, 2023). These correlations offer crucial insights into the underlying physical processes driving the variability in AGNs.

Among the different techniques used to characterize the variability of AGNs, it has become increasingly popular in recent years to model the light curve of AGNs using stochastic processes (e.g., B. C. Kelly et al. 2009; S. Kozłowski et al. 2010; B. C. Kelly et al. 2014). This method solves the problems of sampling and window effects that arise in frequency domain time series analysis, which is particularly related to the light curves of AGNs. The damped random walk (DRW) model has become the simplest Gaussian random process model, which can well fit the light curve of AGNs (e.g., B. C. Kelly et al. 2009; S. Kozłowski et al. 2010; C. L. MacLeod et al. 2010). Generally, the DRW model can successfully fit the long-term variability of the AGN accretion disk. This stochastic process model has been shown to provide a powerful tool for extracting information from AGN variability (e.g., V. P. Kasliwal et al. 2017; C. J. Burke et al. 2021; H. Zhang et al. 2022, 2023). In the DRW model, power spectrum density (PSD) is described by the $f^{-2}$ power law at the high-frequency end and transforms into white noise at the low-frequency end. The transition frequency $f_0$ corresponds to the damping timescale $\tau_{\rm DRW}$, $f_0 = 1/(2\pi\tau_{\rm DRW})$. Therefore, the damping timescale describes the characteristic timescale of optical variability. Early studies of the variability of AGNs have hinted at this characteristic timescale of variability and its possible dependence on physical properties of AGNs, such as black hole mass (e.g., S. Collier & B. M. Peterson 2001; B. C. Kelly et al. 2009). Recently, C. J. Burke et al. (2021) used the DRW model to measure the damping timescale of AGNs with high-quality optical light curves over a large dynamic range of black hole masses. They found a strong positive correlation between the damping timescales and black hole mass, which extends to the stellar mass range with optical variability measured for nova-like accreting white dwarfs.

Although many studies have investigated the properties of optical variability of AGNs, we do not fully understand the mechanisms driving such variability. In particular, it is unclear how the physical properties of the central engine (such as jet power, black hole mass, and Eddington ratio) are related to the variability properties of the system (such as characteristic timescales). According to the theory of jet formation (e.g., R. D. Blandford & R. L. Znajek 1977; R. D. Blandford & D. G. Payne 1982), the central engine is related to the magnetic field and the spin of the black hole. At the same time, some authors have proposed the idea of the magnetic origin of AGNs variabilities (e.g., E. P. Velikhov 1959; S. Chandrasekhar 1961; S. A. Balbus & J. F. Hawley 1991; A. R. King et al. 2004; M. Mayer & J. E. Pringle 2006; A. Janiuk & B. Czerny 2007; J. D. Hogg & C. S. Reynolds 2016). Recently, H. Zhou & D. Lai (2024) discovered that the UV/optical variability of AGNs is interpreted using a quasi-periodic large-scale magnetic field through a one-dimensional thin accretion disk model. These results may imply that the optical variability of AGNs is related to the magnetic field of the jet. However, there are no studies investigating the relationship between the optical variability of AGNs and the jet magnetic field and black hole spin.

In this paper, we study the relation between AGN variability and physical properties of SMBH, such as the relation between characteristic timescales and the magnetic field of the jet, black hole spin, and black hole mass. The paper is organized as follows. We describe the sample and method in Section 2. In Section 3, we describe the results. Section 4 presents discussions. In Section 5, we show the conclusions of the statistical analysis. Throughout the paper, we adopt a cosmology with $H_0 = 70 \text{ km s}^{-1} \text{ Mpc}^{-1}$, $\Omega_m = 0.3$, and $\Omega_\Lambda = 0.7$.

## 2. The Sample and Method

### 2.1. The Sample

We consider the sample of AGNs with jet magnetic fields, black hole mass, broad-line region luminosity, and redshifts. We consider the sample of A. B. Pushkarev et al. (2012). A. B. Pushkarev et al. (2012) used multi-frequency Very Long Baseline Array observations to estimate the jet magnetic fields of 104 AGNs. We get the light curves of 100 AGNs in the DR23 ZTF public data release[8] (F. J. Masci et al. 2019). We only use the g-band light curve because the g-band magnitude is rarely influenced by the host galaxy compared with the r-band magnitude. To obtain a reliable light curve, we use the same strategy as V. Negi et al. (2022). We consider the following criteria: (i) The number of observations for each source was greater than 30. (ii) We only take the light curve corresponding to the observation ID with the maximum number of data points due to combining light curves from different IDs for the same source could produce spurious variability (J. van Roestel et al. 2021; V. Negi et al. 2022). (iii) The ZTF DR23 also provides a quality score of the observed data obtained in good observing conditions,[9] as catflags score = 0. (iv) We rule out any light curve that did not have sufficient signal-to-noise ratio or duration to robustly constrain the damping timescale (C. J. Burke et al. 2021). On the whole, 90 AGNs passed the above criteria.

In order to estimate the spin of black holes, we consider that these sources have reliable black hole mass and broad-line region luminosity. We consider the samples of V. S. Paliya et al. (2021) and M. Zamaninasab et al. (2014). From the catalog of V. S. Paliya et al. (2021) and M. Zamaninasab et al. (2014), we get 90 sources with black hole mass and broad-line region luminosity. At the same time, we get 150 MHz radio flux from the TIFR Giant Metrewave Radio Telescope Sky Survey (H. T. Intema et al. 2017) and 1.4 GHz radio flux from the NRAO VLA Sky Survey (NVSS; J. J. Condon et al. 1998).

### 2.2. The DRW Model

The DRW process is described by a first-order stochastic differential equation (see details in B. C. Kelly et al. 2009). The DRW model is the simplest case of a family of continuous autoregressive with moving average (CARMA) models for Gaussian processes (B. C. Kelly et al. 2014; C. J. Burke et al. 2021). Unlike most high-order CARMA models, the DRW model interprets curvature in the PSD as being due to the damping of power beyond a characteristic timescale (C. J. Burke et al. 2021). For this reason, we use the DRW model as a benchmark model to measure the characteristic break timescale/frequency in the PSD, but we acknowledge that the actual variability process may be more complex than a DRW. We used the *celerite* package to perform Gaussian process regression and fit the light curve with the DRW model, which is the simplest celerite term. Many authors have successfully used the celerite package to study the AGN variability (e.g., C. J. Burke et al. 2021; Z. Stone et al. 2022; H. Zhang et al. 2022, 2023; H. Zhou & D. Lai 2024). The kernel function of the DRW model plus the white noise term in the celerite can be expressed as follows:

$$k(t_{nm}) = 2\sigma_{\text{DRW}}^2 \exp(-t_{nm}/\tau_{\text{DRW}}) + \sigma_n^2 \delta_{nm}, \quad (1)$$

where $\sigma_{\text{DRW}}$ is the amplitude term, $t_{nm}$ is the time lag between measurements $m$ and $n$, and $\tau_{\text{DRW}}$ is the damping timescale. The $\sigma_n$ is the excess white noise amplitude, and $\delta_{nm}$ is the Kronecker $\delta$ function. We use the DRW model to fit the light curve of 90 AGNs based on the *taufit*[10] code (C. J. Burke et al. 2021). The example is shown in Figure 1. To obtain reliable measurements of the damping timescale, we use the following criteria: (1) the damping timescale $\tau_{\text{DRW}} < 0.1 \times$ baseline; (2) the damping timescale $\tau_{\text{DRW}} >$ mean cadence; (3) signal-to-noise: $\sigma_{\text{DRW}} > \sigma_n^2 + \bar{d}y$ (C. J. Burke et al. 2021; H. Zhang et al. 2022), where $\bar{d}y$ is the mean size of the quoted light curve uncertainties. Finally, it is found that the damping timescales for the 41 sources are reliable. We also find that the 41 sources are blazars, namely jetted AGNs.

### 2.3. The Black Hole Spin and Jet Kinetic Power

R. A. Daly (2019) used the following method to calculate the spin of the black hole,

$$\frac{f(j)}{f_{\text{max}}} = \left(\frac{L_j}{g_j L_{\text{Edd}}}\right)\left(\frac{L_{\text{bol}}}{g_{\text{bol}} L_{\text{Edd}}}\right)^{-A}$$

$$j = \frac{2\sqrt{f(j)/f_{\text{max}}}}{f(j)/f_{\text{max}} + 1}, \quad (2)$$

where $L_j$ is the beam power, $L_{\text{Edd}}$ is the Eddington luminosity ($L_{\text{Edd}} = 1.3 \times 10^{38}(M_{\text{BH}}/M_\odot)$), and $g_j = 0.1$, $g_{\text{bol}} = 1$, and $A = 0.43$ are adopted (R. A. Daly 2019). The $L_{\text{bol}}$ is the bolometric luminosity (H. Netzer 1990), $L_{\text{bol}} = 10 L_{\text{BLR}}$. The black hole spin obtained by this method is consistent with the black hole spin obtained by other methods (R. A. Daly 2019), such as the X-ray reflection method (C. S. Reynolds 2014). J. M. Miller et al. (2009) used the X-ray reflection method to obtain the black hole spin of GX 339-4 is $0.94 \pm 0.02$, and the black hole spin is $0.92 \pm 0.06$ using the method of R. A. Daly (2019). Due to the limitations of observation, it is very difficult to directly measure the black hole spin of large samples using the X-ray reflection method. Therefore, we use the method of R. A. Daly (2019) to measure the black hole spin of our sample. Our sample is shown in Table 1.

The beam power can be calculated by the following formula (C. J. Willott et al. 1999),

$$L_j \approx 1.7 \times 10^{45} f^{3/2} \left(\frac{L_{151}}{10^{44} \text{ erg s}^{-1}}\right)^{6/7} \text{ erg s}^{-1}, \quad (3)$$

---

[8] https://www.ztf.caltech.edu
[9] https://irsa.ipac.caltech.edu/data/ZTF/docs/releases/dr23
[10] https://github.com/burke86/taufit

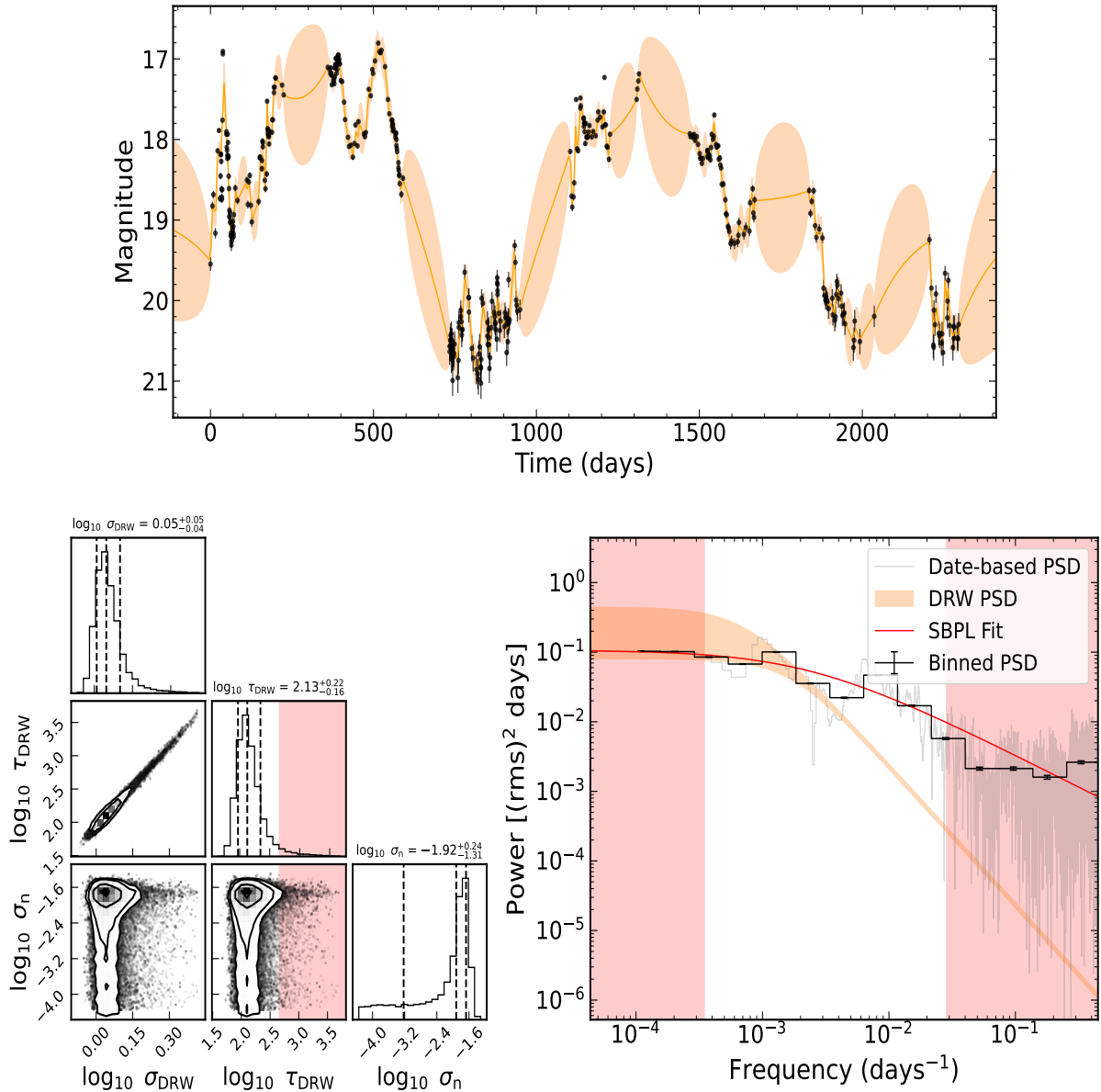

**Figure 1.** An example of fitting the $zg$-band light curve of B0215+015 with the DRW model using the fast Gaussian process solver celerite. The top panel shows the light curve of B0215+015, and the predicted light curve from the DRW model using the best-fitting, maximum likelihood parameters. The orange line is the best-fitting DRW model with $1\sigma$ uncertainty (orange shaded area). The lower-left panel shows the probability distributions for the fitted DRW parameters. The lower-right shows the normalized PSD and binned PSD with $1\sigma$ uncertainties. The orange shaded area is the model PSD with $1\sigma$ uncertainties. The gray line represents the Lomb–Scargle periodogram, while the black line denotes the binned Lomb–Scargle periodogram. The binned Lomb–Scargle periodogram was further fitted to a broken power law, as indicated by the red line. The red shaded regions indicate periods longer than 20% of the light curve duration (as shown in the lower-left and lower-right panels) and shorter than the mean cadence (as displayed in the lower-right panel). The discrepancies between the Lomb–Scargle periodogram and the model power spectral density (PSD) can be attributed to several factors: the inherent challenges in accurately estimating the PSD using Fourier-based methods on irregularly sampled light curves, the influence of flux measurement uncertainties on periodogram accuracy, and potential departures from the damped random walk (DRW) model (Z. Stone et al. 2022).

where $L_{151}$ is the 151 MHz radio luminosity in units of erg s$^{-1}$, $f = 1$ is adopted (X. Cao 2003). The 151 MHz radio luminosity is estimated using $L_\nu = 4\pi d_L^2 S_\nu$, $d_L(z) = \frac{c}{H_0}(1 + z) \int_0^z [\Omega_\Lambda + \Omega_m(1 + z')^3]^{-1/2} dz'$, where $d_L$ is the luminosity distance (T. M. Venters et al. 2009). We make a K-correction for the observed flux using $S_\nu = S_\nu^{\rm obs}(1 + z)^{\alpha-1}$ and $\alpha = 0.8$ (P. Cassaro et al. 1999).

We use the following formula to calculate the jet kinetic power (K. W. Cavagnolo et al. 2010),

$$\log P_{\rm jet} = 0.75(\pm 0.14) \log P_{1.4} + 1.91(\pm 0.18) \quad (4)$$

where $P_{1.4}$ is the 1.4 GHz radio luminosity in units of $10^{40}$ erg s$^{-1}$, and $P_{\rm jet}$ is in units of $10^{42}$ erg s$^{-1}$. The scatter for this relation is $\sigma = 0.78$ dex.

## 3. Results

### 3.1. Relations between Damping Timescale and Jet Magnetic Field, Black Hole Spin, and Jet Kinetic Power

We use linear regression to analyze the relationship between damping timescale ($\tau_{\rm DRW}$) and jet magnetic field, and black hole spin of AGNs, respectively. The relation between damping timescale and jet magnetic field (a), black hole spin

**Table 1**
The Sample of Jetted AGNs

| Name (1) | R.A. (2) | Decl. (3) | $z_*$ (4) | $B_{1\ pc}$ (5) | $S_{151}$ (6) | $\log j$ (7) | $S_{1.4}$ (8) | $\log M_{BH}$ (9) | $\log L_{BLR}$ (10) | $\log \tau_{DRW}$ (11) | SNR (12) |
|---|---|---|---|---|---|---|---|---|---|---|---|
| QSO B0106+013 | 17.1615 | 1.5834 | 2.099 | 0.79 | 5179.2 ± 518 | −0.154 | 2620.7 ± 78.6 | 9.64 ± 0.11 | 45.62 ± 0.03 | $2.41^{0.37}_{-0.21}$ | 21.3 |
| QSO B0149+218 | 28.0752 | 22.1188 | 1.32 | 1.69 | 2959.2 ± 296 | −0.072 | 1041.1 ± 31.2 | 8.45 ± 0.06 | 45.17 ± 0.02 | $1.92^{0.20}_{-0.15}$ | 5.8 |
| QSO B0133+476 | 24.2441 | 47.8581 | 0.859 | 0.77 | 1027.5 ± 102.9 | −0.23 | 1137.5 ± 34.1 | 8.33 ± 0.29 | 44.6 ± 0.05 | $2.09^{0.22}_{-0.15}$ | 12.19 |
| QSO 0215+015 | 34.454 | 1.7471 | 1.715 | 1.41 | 1453.4 ± 145.5 | −0.182 | 750.2 ± 22.5 | 9.22 ± 0.21 | 44.85 ± 0.04 | $2.13^{0.22}_{-0.16}$ | 15.8 |
| QSO 0234+285 | 39.4684 | 28.8025 | 1.207 | 1.71 | 2327.9 ± 232.9 | −0.312 | 2196.9 ± 65.9 | 9.13 ± 0.25 | 45.4 ± 0.07 | $2.08^{0.23}_{-0.15}$ | 28.4 |
| QSO B0336-019 | 54.8789 | −1.7766 | 0.852 | 0.92 | 955.3 ± 95.7 | −0.418 | 2423.6 ± 72.7 | 8.7 ± 0.32 | 45.03 ± 0.07 | $1.89^{0.18}_{-0.14}$ | 18.27 |
| QSO B0420-014 | 65.8158 | −1.3425 | 0.914 | 1.41 | 951 ± 95.5 | −0.244 | 2725.8 ± 81.8 | 8.4 ± 0.23 | 44.65 ± 0.05 | $1.79^{0.14}_{-0.11}$ | 21.82 |
| QSO B0836+710 | 130.3515 | 70.895 | 2.218 | 1.93 | 4429.7 ± 443 | −0.135 | 3823.1 ± 114.7 | 8.94 ± 0.3 | 46.4 ± 0.07 | $2.48^{0.48}_{-0.25}$ | 16.43 |
| QSO B0528+134 | 82.7351 | 13.532 | 2.07 | 1.6 | 1601.2 ± 160.3 | 0 | 1556.2 ± 46.7 | 8.17 ± 0.13 | 44.73 ± 0.05 | $2.03^{0.38}_{-0.30}$ | 5.4 |
| B0730+504 | 113.4688 | 50.3692 | 0.72 | 1.47 | 1052.6 ± 105.4 | −0.233 | 769.6 ± 23.1 | 8.26 ± 0.13 | 44.37 ± 0.05 | $1.58^{0.11}_{-0.10}$ | 15.04 |
| QSO B0748+126 | 117.7169 | 12.518 | 0.889 | 0.84 | 1707.6 ± 170.9 | −0.549 | 1452.8 ± 43.6 | 9.09 ± 0.29 | 45.75 ± 0.05 | $1.5^{0.10}_{-0.09}$ | 16.2 |
| QSO B0804+499 | 122.1653 | 49.8435 | 1.436 | 0.48 | 450.9 ± 45.9 | −0.3 | 1114.5 ± 33.4 | 8.51 ± 0.13 | 45.09 ± 0.02 | $0.97^{0.12}_{-0.11}$ | 11.97 |
| QSO B1849+670 | 282.317 | 67.0949 | 0.657 | 0.52 | 990.5 ± 99.7 | −0.289 | 517.4 ± 15.5 | 8.3 ± 0.61 | 44.39 ± 0.06 | $1.77^{0.22}_{-0.20}$ | 2.25 |
| QSO B1015+359 | 154.5457 | 35.711 | 1.226 | 0.92 | 532.1 ± 53.8 | −0.461 | 615.2 ± 18.5 | 8.82 ± 0.02 | 45.33 ± 0 | $2.16^{0.26}_{-0.24}$ | 1.68 |
| QSO B1127-145 | 172.5294 | −14.8243 | 1.184 | 0.84 | 3293 ± 329.6 | −0.348 | 5622.2 ± 198.5 | 9.26 ± 0.29 | 45.68 ± 0.07 | $2.12^{0.34}_{-0.22}$ | 13.85 |
| QSO B1156+295 | 179.8826 | 29.2455 | 0.73 | 1.17 | 4915.2 ± 491.6 | −0.214 | 2030.8 ± 71.8 | 8.98 ± 0.07 | 44.67 ± 0.03 | $1.93^{0.16}_{-0.12}$ | 59.94 |
| QSO B1219+044 | 185.5939 | 4.2211 | 0.965 | 0.81 | 1544.8 ± 154.8 | −0.176 | 800.3 ± 24 | 8.25 ± 0.1 | 44.98 ± 0.02 | $1.31^{0.11}_{-0.10}$ | 7.89 |
| QSO B1253-055 | 194.0465 | −5.7893 | 0.536 | 0.42 | 20732 ± 2073.3 | −0.026 | 9711.2 ± 291.3 | 8.63 ± 0.38 | 44.28 ± 0.05 | $1.7^{0.09}_{-0.08}$ | 4.87 |
| QSO B1308+326 | 197.6194 | 32.3455 | 0.997 | 0.96 | 2240.3 ± 224.3 | −0.203 | 1686.6 ± 50.6 | 8.72 ± 0.05 | 44.91 ± 0.01 | $2.37^{0.35}_{-0.20}$ | 35.5 |
| QSO B1334-127 | 204.4158 | −12.9569 | 0.539 | 1.23 | 2095.1 ± 209.6 | −0.311 | 2676.3 ± 80.3 | 8.62 ± 0.53 | 44.33 ± 0.14 | $1.65^{0.13}_{-0.11}$ | 17.71 |
| QSO B1510-089 | 228.2106 | −9.1 | 0.36 | 0.73 | 2475.2 ± 248 | −0.439 | 2700.9 ± 81 | 8.32 ± 0.13 | 44.72 ± 0.04 | $1.51^{0.14}_{-0.11}$ | 27.28 |
| QSO B1508-055 | 227.7233 | −5.7187 | 1.191 | 1.52 | 9489.2 ± 949 | −0.016 | 3569.3 ± 107.1 | 8.53 ± 0.34 | 45.22 ± 0.1 | $2.34^{0.32}_{-0.26}$ | 1.6 |
| QSO B1502+106 | 226.1041 | 10.4942 | 1.839 | 0.69 | 1459 ± 146.1 | −0.23 | 1774.2 ± 53.2 | 9.13 ± 0.06 | 45.4 ± 0.01 | $2.18^{0.22}_{-0.15}$ | 17.81 |
| QSO B1532+016 | 233.7186 | 1.5178 | 1.42 | 1.14 | 823.4 ± 82.6 | −0.318 | 1320.4 ± 39.6 | 8.93 ± 0.06 | 45.13 ± 0.03 | $1.44^{0.14}_{-0.12}$ | 4.95 |
| QSO B1546+027 | 237.3727 | 2.617 | 0.414 | 0.32 | 515.7 ± 51.9 | −0.742 | 835.3 ± 29.5 | 8.45 ± 0.01 | 44.93 ± 0 | $1.16^{0.11}_{-0.10}$ | 2.49 |
| QSO B1538+149 | 235.2062 | 14.7961 | 0.605 | 0.49 | 4395.9 ± 439.7 | −0.002 | 1386.8 ± 41.6 | 8.23 ± 0.26 | 43.18 ± 0.18 | $1.86^{0.14}_{-0.11}$ | 23.96 |
| QSO B1606+106 | 242.1925 | 10.4855 | 1.226 | 0.79 | 784.1 ± 78.6 | −0.328 | 1392 ± 41.8 | 8.74 ± 0.06 | 45.09 ± 0.02 | $1.34^{0.08}_{-0.07}$ | 16.82 |
| QSO B1641+399 | 250.745 | 39.8103 | 0.593 | 1.2 | 10011.9 ± 1001.3 | −0.146 | 7098.6 ± 213 | 8.66 ± 0.02 | 44.86 ± 0 | $1.72^{0.11}_{-0.09}$ | 28.68 |
| QSO B1730-130 | 263.2613 | −13.0804 | 0.902 | 1.69 | 7650.6 ± 765.2 | −0.03 | 5990.2 ± 179.7 | 8.45 ± 0.21 | 44.77 ± 0.13 | $2.29^{0.50}_{-0.32}$ | 3.8 |
| QSO B1726+455 | 261.8652 | 45.511 | 0.717 | 0.28 | 395.6 ± 40.1 | −0.525 | 913.9 ± 27.4 | 8.25 ± 0.3 | 45.04 ± 0.06 | $1.01^{0.05}_{-0.05}$ | 31.46 |
| QSO B1749+096 | 267.8867 | 9.6502 | 0.322 | 0.33 | 594.1 ± 59.8 | −0.315 | 622.6 ± 18.7 | 7.98 ± 0.51 | 43.08 ± 0.18 | $1.27^{0.07}_{-0.06}$ | 44.25 |
| QSO B2131-021 | 323.543 | −1.8881 | 1.285 | 1.02 | 2320.8 ± 232.3 | −0.014 | 1689.5 ± 59.7 | 8.75 ± 0.38 | 43.83 ± 0.19 | $1.34^{0.09}_{-0.08}$ | 26.71 |
| QSO B2121+053 | 320.9355 | 5.5895 | 1.941 | 1.43 | 429.9 ± 44.5 | −0.576 | 793.5 ± 23.8 | 9.33 ± 0.51 | 45.96 ± 0.11 | $1.82^{0.14}_{-0.11}$ | 10.21 |
| QSO B2113+293 | 318.8726 | 29.5607 | 1.514 | 1.16 | 445.9 ± 44.8 | −0.302 | 755.2 ± 22.7 | 8.94 ± 0.22 | 44.63 ± 0.09 | $1.99^{0.16}_{-012}$ | 11.56 |
| QSO B2209+236 | 333.0249 | 23.9279 | 1.125 | 0.39 | 198.2 ± 20.3 | −0.492 | 556.5 ± 16.7 | 8.69 ± 0.04 | 44.62 ± 0.05 | $1.69^{0.12}_{-0.10}$ | 11.28 |
| QSO B2227-088 | 337.417 | −8.5485 | 1.56 | 1.44 | 545.3 ± 55.1 | −0.404 | 968.1 ± 29 | 8.86 ± 0.03 | 45.5 ± 0.01 | $2.02^{0.21}_{-0.16}$ | 12.96 |
| QSO B2223-052 | 336.4469 | −4.9504 | 1.404 | 1.47 | 20306.6 ± 2030.7 | −0.001 | 7410.6 ± 261.6 | 8.66 ± 0.19 | 45.6 ± 0.03 | $2.06^{0.20}_{-0.15}$ | 11.66 |
| QSO B2230+114 | 338.1517 | 11.7308 | 1.037 | 2.12 | 5903 ± 590.5 | −0.192 | 7201.5 ± 216 | 9.09 ± 0.32 | 45.27 ± 0.19 | $2.22^{0.17}_{-0.13}$ | 4.98 |
| QSO B2251+158 | 343.4906 | 16.1482 | 0.859 | 1.13 | 8343.8 ± 834.7 | −0.189 | 12656.8 ± 379.7 | 8.95 ± 0.24 | 45.35 ± 0.07 | $1.49^{0.09}_{-0.08}$ | 31.45 |
| QSO B2345-167 | 357.0109 | −16.52 | 0.576 | 0.9 | 3207.4 ± 321.1 | −0.179 | 2641.5 ± 79.2 | 8.44 ± 0.25 | 44.32 ± 0.09 | $1.51^{0.12}_{-0.10}$ | 38.86 |
| QSO B2351+456 | 358.5903 | 45.8845 | 1.986 | 1.72 | 1813.2 ± 181.5 | −0.145 | 1872.3 ± 56.2 | 9.2 ± 0.11 | 45.13 ± 0.04 | $2.17^{0.56}_{-0.42}$ | 3.37 |

**Note.** Column (1): name; Column (2): the R.A. in decimal degrees; Column (3): (delineation) in decimal degrees; Column (4): redshift; Column (5): magnetic field strength one parsec away from the jet base (in units Gauss); Column (6): 151 MHz flux in units mjy; Column (7): black hole spin; Column (8): 1.4 GHz radio flux in units mjy; Column (9): black hole mass; Column (10): the broad-line luminosity; Column (11): the optical variability damping timescale in units days; Column (12): the signal-to-noise.

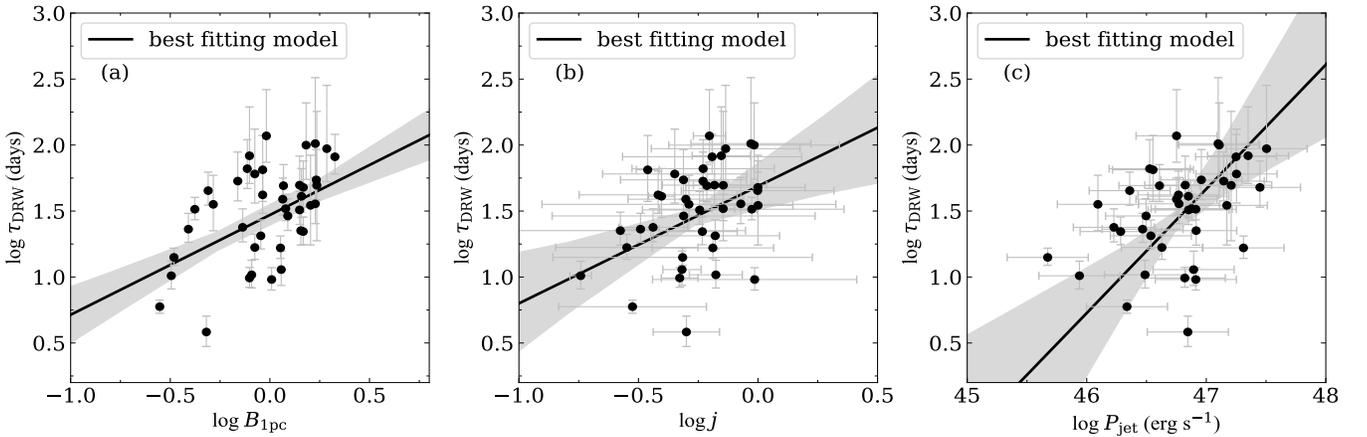

**Figure 2.** The relations between the damping timescale in the rest frame ($\tau_{\mathrm{DRW}}$) and other physical parameters. (a) Relation between damping timescale in the rest frame ($\tau_{\mathrm{DRW}}$) and the magnetic field of the jets. (b) Relation between damping timescale in the rest frame ($\tau_{\mathrm{DRW}}$) and black hole spin. (c) Relation between damping timescale in the rest frame ($\tau_{\mathrm{DRW}}$) and jet kinetic power. Shaded gray colored areas correspond to $1\sigma$ confidence bands. The black line is the best fit.

(b) and jet kinetic power (c) for our sample are shown in Figure 2. A fit to the relation between the two variables using linmix[11] (B. C. Kelly 2007). There is a significant correlation between damping timescale and jet magnetic field for our sample (coefficient of correlation $r = 0.54$, and significance level $P = 0.0005$; $P < 0.05$ represents a significant correlation at 95% confidence level). The best-fitting model of the relation between damping timescale and jet magnetic field is:

$$\log \tau_{\mathrm{DRW}} = 0.76(\pm 0.22) \log B_{1\,\mathrm{pc}} + 1.47(\pm 0.05). \quad (5)$$

where $B_{1\,\mathrm{pc}}$ is the jet magnetic field. The $1\sigma$ intrinsic scatter of the best-fitting model is $0.08 \pm 0.02$. Our results may suggest that the optical variability of our sample is mainly driven by the magnetic field of the jet.

The relation between the damping timescale and black hole spin for our sample is shown in the middle panel of Figure 2. We find a significant correlation between damping timescale and black hole spin for our sample ($r = 0.45$, $P = 0.0012$). The best-fitting model of the relation between damping timescale and black hole spin for our sample is:

$$\log \tau_{\mathrm{DRW}} = 0.92(\pm 0.55) \log j + 1.70(\pm 0.17). \quad (6)$$

where $j$ is the black hole spin. The $1\sigma$ intrinsic scatter of this relation is $0.09 \pm 0.03$. According to the theory of jet formation (R. D. Blandford & R. L. Znajek 1977), the spin of the black hole enhances the relativistic jet. Our results may suggest that the optical variability of AGNs is related to relativistic jets.

The relations between damping timescale and jet kinetic power (c) for our sample are shown in the right panel of Figure 2. There is a significant correlation between damping timescale and jet kinetic power for our sample ($r = 0.64$, $P < 0.0001$). The best-fitting model of the relation between damping timescale and jet kinetic power is:

$$\log \tau_{\mathrm{DRW}} = 1.00(\pm 0.72) \log P_{\mathrm{jet}} - 47.00(\pm 34.00). \quad (7)$$

The $1\sigma$ intrinsic scatter of this relation is $0.06 \pm 0.03$. According to the theory of jet formation (R. D. Blandford & R. L. Znajek 1977), the power of the jet depends on the magnetic field strength and the spin of the black hole. We have found a significant relationship between the jet magnetic field, the spin of the black hole, and the damping timescale, respectively. Therefore, the relation between the damping timescale and jet kinetic power further implies a close relationship between the optical variability of AGNs and jets.

---

[11] https://linmix.readthedocs.io/en/latest/

However, it is crucial to acknowledge that the influence of jet power is entangled with two other significant factors: the spin of the black hole and the jet's magnetic field. This degeneracy makes it challenging to precisely isolate the effect of jet power on the observed phenomena. To address this issue and gain a clearer understanding of the relationship, we employ the method of partial correlation to re-examine the connection between the damping timescale and jet power. By statistically controlling for the impact of the black hole spin, we find that there exists a relatively weak correlation between the damping timescale and jet power, with a correlation coefficient of $r = 0.30$ and a corresponding $P$ value of $P = 0.06$. Furthermore, when we accounted for the influence of the jet magnetic field, the analysis again revealed a weak correlation. In this case, the correlation coefficient was $r = 0.16$, and the $P$ value was $P = 0.33$. These results suggest that while there may be some association between the damping timescale and jet power, the relationship is not highly robust, and other factors likely play important roles in determining the damping timescale.

### 3.2. Relations between Damping Timescale and Black Hole Mass

The relation between the damping timescale and the mass of black holes for our sample is shown in Figure 3. We find a significant correlation between damping timescale and the mass of black holes for our sample ($r = 0.51$, $P < 0.0001$). The best-fitting model of the relation between damping timescale and black hole mass is:

$$\log \tau_{\mathrm{DRW}} = 0.52(\pm 0.21) \log M_{\mathrm{BH}}/M_\odot - 3.12(\pm 1.90). \quad (8)$$

The $1\sigma$ intrinsic scatter of this relation is $0.08 \pm 0.02$.

### 4. Discussion

#### 4.1. The Correlations between Variability Damping Timescale and Jet's Physical Parameters

The optical variability of AGNs may come from relativistic jets (e.g., B. C. Kelly et al. 2009; S. Rakshit & C. S. Stalin 2017;

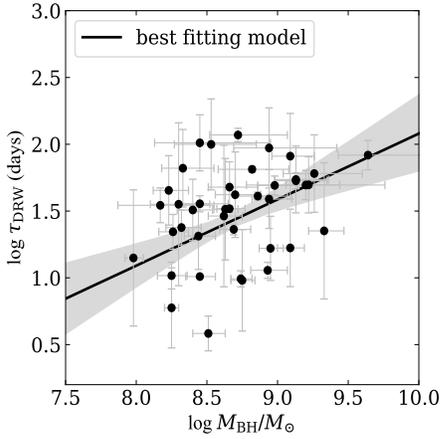

**Figure 3.** Relations between damping timescale in the rest frame ($\tau_{\rm DRW}$) and black hole mass for our sample. Shaded gray colored areas correspond to $1\sigma$ confidence bands. The black line is the best fit.

D. Xiong et al. 2025). Therefore, we investigated the relationship between the variability of AGNs and physical parameters of jets, such as the magnetic field of jets, the spin of black holes, and the power of jets. We find a significant correlation between the damping timescale and the magnetic field of jets for our sample. M. Böttcher et al. (2010) studied the optical spectral variability of the blazars 1ES 1011+496 and found that the optical variability of 1ES 1011+496 is primarily driven by changes in the magnetic field. A. A. Abdo et al. (2010) studied the gamma-ray flare of 3C 279 and found that there is a relation between the gamma-ray flare of 3C 279 and a highly ordered jet magnetic field. Some authors have suggested that the intrinsic fluctuations of turbulence driven by the magnetorotational instability in the accretion disk may lead to optical variability of AGNs (E. P. Velikhov 1959; S. Chandrasekhar 1961; S. A. Balbus & J. F. Hawley 1991). Recently, H. Zhou & D. Lai (2024) proposed that quasi-periodic large-scale magnetic fields may explain the optical variability of AGNs.

The changes in these physical parameters (e.g., magnetic field, particle acceleration rate, cooling rate, escape rate) may cause the generation of variability in the jet-emitting region. The characteristic timescales of variability in AGNs correspond to these timescales (e.g., light-crossing timescale, cooling timescale, acceleration timescale, escape timescale). The escape timescale of diffusion can be expressed by (R. Xue et al. 2019)

$$t_{\rm esc} = \frac{R^2}{4D} = \frac{3eBR^2}{4\alpha \gamma_e m_e c^3}$$
$$= 1.2 {\rm yr} \left(\frac{10^6}{\gamma_e}\right)\left(\frac{B}{10^{-3}\,{\rm G}}\right)\left(\frac{10^5}{\alpha}\right)\left(\frac{R}{5\times 10^{17}\,{\rm cm}}\right), \quad (9)$$

where $D$ is the diffusion coefficient, $R$ is the radius of the emission zone, $\alpha$ is the ratio of the mean magnetic field energy density to the turbulent magnetic field energy density, and $\gamma_e$ is the characteristic random Lorentz factor of electrons. The relation between the jet magnetic field and the escape timescale is $t_{\rm esc} \propto B^{1.0}$. We find that the slope of the relation between the damping timescale and the jet magnetic field for our sample is $\tau_{\rm DRW} \propto B_{\rm 1\,pc}^{0.76\pm 0.22}$. Our result is consistent with the escape times scale of diffusion within the range of error.

### 4.2. The Correlations between Variability Damnping Timescale and Black Hole Mass

Some authors have found a significant correlation between the damping timescale and the mass of black holes for nonjetted AGNs (e.g., B. C. Kelly et al. 2009; C. L. MacLeod et al. 2010; C. J. Burke et al. 2021; K. L. Suberlak et al. 2021; Z.-B. Su et al. 2024; H. Zhou & D. Lai 2024). B. C. Kelly et al. (2009) found that the relation between damping timescale and black hole mass for 70 quasars is $\log \tau_{\rm DRW} \sim (0.56 \pm 0.14)\log M_{\rm BH}$. S. Kozłowski (2016) found that the relation between damping timescale and black hole mass for AGNs from Stripe 82 of the Sloan Digital Sky Survey is $\tau_{\rm DRW} \propto M_{\rm BH}^{0.38\pm 0.15}$. C. J. Burke et al. (2021) found that the relation between damping timescale and black hole mass for 67 nonjetted AGNs is $\log \tau_{\rm DRW} \sim 0.38^{+0.05}_{-0.04} \log M_{\rm BH}$. H. Zhang et al. (2022) found that the relation between damping timescale and black hole mass for 23 jetted AGNs is $\log \tau_{\rm DRW} \sim (0.43 \pm 0.04)\log M_{\rm BH}$. H. Zhang et al. (2024) found that the relation between damping timescale and black hole mass for 34 blazars and seven microquasars is $\log \tau_{\rm DRW} \sim 0.57^{+0.02}_{-0.02} \log M_{\rm BH}$. Z.-B. Su et al. (2024) found a new scaling relation between the damping timescale and black hole mass for the intermediate-mass black hole and SMBH by using the structure function method: $\tau \propto M_{\rm BH}^{0.6-0.8}$. We find that the slope of the relation between the damping timescale and black hole mass for our jetted AGNs is given by $\log \tau_{\rm DRW} \sim 0.52(\pm 0.21)\log M_{\rm BH}$. The slopes describing the relation between damping timescale and black hole mass for jetted and nonjetted AGNs are similar within the error range. These results may imply that the origin of optical variability for jetted AGNs and nonjetted AGNs is potentially similar.

Within the framework of a standard accretion disk model, the thermal timescale is proportional to the black hole mass (C. J. Burke et al. 2021), with $\tau \propto M_{\rm BH}^{0.50}$. The slope observed in our sample aligns closely with this theoretical relation between thermal timescale and black hole mass. Furthermore, we observe a significant correlation between the damping timescale and the magnetic field strength of the jets in our sample. The magnetic field strength of the jet influences the magnetic field within the accretion disk, which in turn drives MRI fluctuations occurring on thermal timescales. These MRI-induced fluctuations are believed to be responsible for the observed variability in the light curves emitted by the disk (A. Secunda et al. 2024; H. Zhou & D. Lai 2024; A. Secunda et al. 2025). Our results may provide supporting evidence that MRI-driven fluctuations are responsible for intrinsic variability in the light curves emitted by the AGN disk.

### 4.3. The Correlations between Amplitude of Variability and Physical Parameters

In addition to the damping timescale, several studies have investigated the correlation between the amplitude of variability and the physical parameters of AGNs (e.g., B. C. Wilhite et al. 2007, 2008; M. Wold et al. 2007; B. C. Kelly et al. 2009; Y. L. Ai et al. 2010; C. L. MacLeod et al. 2010; B. C. Kelly et al. 2013; T. Simm et al. 2016; S. Rakshit & C. S. Stalin 2017; Z. Li et al. 2018; P. Sánchez-Sáez et al. 2018; M. Laurenti et al. 2020; W. Yu et al. 2022; P. Arévalo et al. 2023). In this study, we employ linear regression analysis to explore the relationship between the structure function at infinity (SF$_\infty \equiv \sqrt{2}\,\sigma_{\rm DRW}$, where $\sigma_{\rm DRW}$ represents the amplitude of variability in the DRW model, C. J. Burke et al. 2021; Z. Stone et al. 2022) and AGN physical parameters. The relationship between variability amplitude and AGN physical properties is illustrated in Figure 4. We find a statistically significant correlation between the amplitude of variability and black hole mass ($r = 0.40$, $P = 0.018$), as well as a significant negative correlation with the Eddington ratio ($r = -0.52$, $P = 0.0029$). These findings align with previous studies: Z. Li et al. (2018) reported a significant negative correlation between variability amplitude and the Eddington ratio based on a sample of $10^5$ quasars. Similarly, Y. L. Ai et al. (2010) observed a positive correlation between variability amplitude and black hole mass, along with an inverse correlation between variability amplitude and Eddington ratio, using a sample of broad-line and narrow-line Seyfert 1 AGNs. S. Rakshit & C. S. Stalin (2017) confirmed these trends within the same AGN subclass. Furthermore, P. Arévalo et al. (2023) and M. Laurenti et al. (2020) independently found that variability amplitude anticorrelates with the Eddington ratio using different AGN samples. Based on the above conclusions, our results further indicate that jetted and nonjetted AGN seem to have similar sources of variability. However, our analysis also shows only weak correlations between variability amplitude and the magnetic field strength of jets ($r = 0.21$, $P = 0.22$) and black hole spin ($r = 0.26$, $P = 0.12$). Additionally, no statistically significant correlation is found between variability amplitude and jet power ($r = 0.09$, $P = 0.75$).

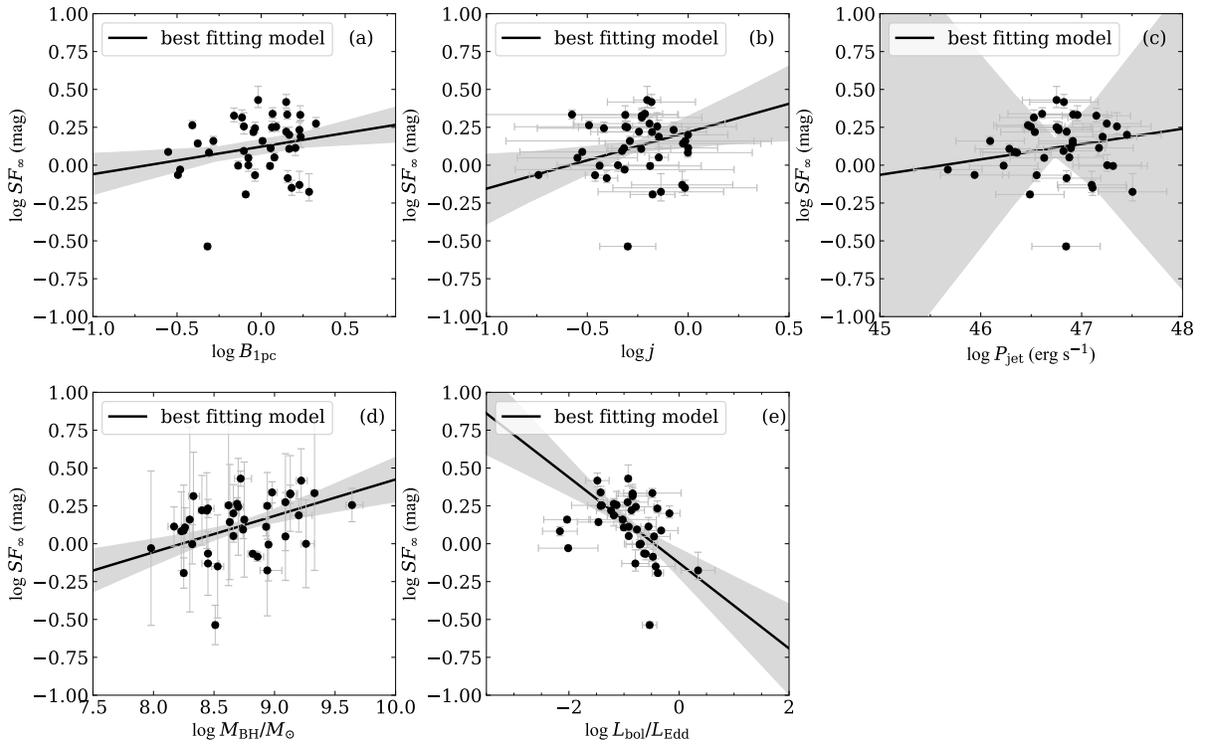

**Figure 4.** Relations between the amplitude of variability and physical parameters for our sample. (a) Relation between amplitude of variability and jet magnetic field. (b) Relation between amplitude of variability and black hole spin. (c) Relation between the amplitude of variability and jet power. (d) Relation between amplitude of variability and black hole mass. (e) Relation between amplitude of variability and Eddington ratio. Shaded gray colored areas correspond to 1σ confidence bands. The black line is the best fit.

Several studies, including B. C. Wilhite et al. (2008), C. L. MacLeod et al. (2010), T. Simm et al. (2016), and S. Rakshit & C. S. Stalin (2017), have explored potential explanations for the observed inverse relationship between variability amplitude and $L_{bol}/L_{Edd}$. One proposed interpretation suggests that $L/L_{Edd}$ may serve as an indicator of AGN age

(e.g., P. Martini & D. P. Schneider 2003; M. Haas 2004; P. F. Hopkins et al. 2005). In this context, objects with lower $L_{bol}/L_{Edd}$ values may be older systems experiencing a reduced supply of accreting material, leading to a more unstable accretion process and consequently greater variability amplitude. Another possible interpretation is that objects with higher $L_{bol}/L_{Edd}$ values are associated with hotter accretion disks, in accordance with classical accretion disk theory (N. I. Shakura & R. A. Sunyaev 1973). Under typical values of black hole mass and accretion rate, the innermost region of the disk is expected to emit predominantly in the far-UV range. Due to its smaller size, this region also exhibits the highest variability amplitude. However, for systems with lower accretion rates, the disk temperature decreases, causing the most variable and innermost region to shift its emission from the UV to the optical wavelength range ($r_\lambda \propto M_{BH}^{2/3}(L_{bol}/L_{Edd})^{1/3}\lambda^{4/3}$).

We also note that our sample has a high accretion rate ($L_{bol}/L_{Edd} \geqslant 5 \times 10^{-3}$, T. Sbarrato et al. 2012), that is, effective accretion. Based on the above discussion, we propose the following physical mechanism: For jetted AGNs with efficient accretion, the observed optical variability may originate within the standard accretion disk apart from the jets. Within this framework, magnetic fields are likely to play a pivotal role across the entire process. The variability originating from the accretion disk may trigger changes in the jet. Due to the coupling between the jets and the accretion disk, changes in the jets may be triggered through the magnetic field. The shock waves generated at the base of the jets may be affected by this magnetic field. Once a shock or turbulence occurs, particles will be accelerated, undergo radiative cooling, and escape, among which the escape process is the dominant factor. S. Rakshit & C. S. Stalin (2017) proposed that, in addition to the accretion disk, jets contribute significantly to the optical variability observed in radio-loud AGNs with high accretion rates.

## 5. Conclusions

In this study, we primarily employ the DRW model to fit the optical light curves of AGNs in order to derive the damping timescale and the amplitude of variability. We investigate the relationships between the damping timescale and several key physical parameters, including the jet magnetic field, black hole spin, jet power, black hole mass, and Eddington ratio. Similarly, we examine the correlations involving the amplitude of variability. Our main findings are summarized as follows:

(i) A significant correlation is observed between the damping timescale and the jet magnetic field and black hole spin for our sample. These results suggest that the optical variability of jetted AGNs is influenced by the jet magnetic field and black hole spin.

(ii) The slope of the relation between the damping timescale and jet magnetic field is consistent with the theoretical prediction within the error range. This consistency provides additional observational support for the hypothesis that AGN variability is linked to the escape timescale of diffusive processes.

(iii) A significant correlation is detected between the damping timescale and black hole mass. Moreover, the slope of this relationship is consistent with that of the thermal timescale versus black hole mass. These findings may indicate that the magnetic field strength of the jet influences the magnetic field in the accretion disk, which drives MRI-induced fluctuations on thermal timescales. These MRI fluctuations then drive variability in light curves emitted by the disk.

(iv) A significant correlation is identified between the amplitude of variability and both black hole mass and Eddington ratio. In contrast, only a weak correlation is observed between the amplitude of variability and the jet magnetic field or black hole spin.

## Acknowledgments

Y.C. is grateful for financial support from the National Natural Science Foundation of China (No. 12203028). Y.C. is grateful for funding for the training Program for talents in Xingdian, Yunnan Province (2081450001). Q.S.G.U. is supported by the National Natural Science Foundation of China (12121003, 12192220, and 12192222). This work is supported by the National Natural Science Foundation of China (11733001, U2031201 and 12433004). X.G. acknowledge the support of National Nature Science Foundation of China (Nos 12303017). This work is also supported by Anhui Provincial Natural Science Foundation project number 2308085QA33. D.R.X. is supported by the NSFC 12473020, Yunnan Province Youth Top Talent Project (YNWR-QNBJ-2020-116) and the CAS Light of West China Program.

## ORCID iDs

Yongyun Chen (陈永云)  https://orcid.org/0000-0001-5895-0189
Qiusheng Gu (顾秋生)  https://orcid.org/0000-0002-3890-3729
Junhui Fan (樊军辉)  https://orcid.org/0000-0002-5929-0968
Dingrong Xiong (熊定荣)  https://orcid.org/0000-0002-6809-9575
Xiaotong Guo (郭晓通)  https://orcid.org/0000-0002-2338-7709
Nan Ding (丁楠)  https://orcid.org/0000-0003-1028-8733
Ting-Feng Yi (易庭丰)  https://orcid.org/0000-0001-8920-0073